# Interactive Timetabling

Tomáš Müller, Roman Barták[*]

Charles University
Department of Theoretical Computer Science
Malostranské náměstí 2/25, Praha 1, Czech Republic
tomas.muller@st.mff.cuni.cz
bartak@kti.mff.cuni.cz

**Abstract.** Timetabling is a typical application of constraint programming whose task is to allocate activities to slots in available resources respecting various constraints like precedence and capacity. In this paper we present a basic concept, a constraint model, and the solving algorithms for interactive timetabling. Interactive timetabling combines automated timetabling (the machine allocates the activities) with user interaction (the user can interfere with the process of timetabling). Because the user can see how the timetabling proceeds and can intervene this process, we believe that such approach is more convenient than full automated timetabling which behaves like a black-box. The contribution of this paper is twofold: we present a generic model to describe timetabling (and scheduling in general) problems and we propose an interactive algorithm for solving such problems.

## Introduction

Timetabling is a form of scheduling where the task is to allocate activities to available slots in resources respecting some constraints. Typically, the activity is described by its duration and by resources required to process the activity (conjunction and disjunction of resources can be used). There are also precedence constraints between the activities stating which activity must proceed before another activity and resource constraints stating when and how many activities can be processed by a given resource. In addition to the above hard constraints, that cannot be violated, there exist many preferences - soft constraints describing users' wishes about the timetable. Constraint programming (CP) is a natural tool for describing and solving such problems and there exist many timetabling systems based on CP [1, 2, 5, 9, 12].

In the current personal computing environment, an interactive behaviour of software is a feature requested by many users. Interactivity manifests itself in two directions: the user can observe what is doing inside the system (this is important for debugging but the final users like the animation of problem solving process as well [personal communication to Helmut Simonis]) and he or she can immediately influence the system. Interactive timetabling is a concept where the system presents to the user how the timetable is being built and the user can interfere with this process by changing the task on flow. Naturally, this requires a special interactive solving algorithm that is building a timetable step by step and that can present a sound (sub-) result anytime. Moreover, such algorithm can be exposed to task change, so it should not start re-scheduling from scratch after such a change but the new solution is built using a former solution. By the task change we mean that the user can schedule the activity manually (and fix it in the slot) and vice versa (remove the activity from the slot to which it is currently allocated). Moreover, he/she can alter some other parameters like activity duration, precedence relation etc.

---

[*] Supported by the Grant Agency of the Czech Republic under the contract no. 201/01/0942.

In this paper we define a generic timetabling problem motivated by school timetabling and we propose an interactive algorithm to solve this problem. It may seem that local search algorithms are best suited for this type of problems but note that typically local search is exploring the space of complete schedules which are not sound and it reduces the number of violations. However, we require a sound (perhaps incomplete) schedule to be presented to the user every time. On the other side backtracking based search, which is a method of extending sound but incomplete solutions to a complete solution, does not support interactivity in the sense of changing the task during search. Therefore we propose a mixture of both approaches that borrows techniques both from backtracking based search (extending a partial solution to a complete solution and value and variable selection criteria) as well as from local search (tabu list to prevent cycling). In some sense, the presented algorithm mimics the behaviour of a human scheduler and therefore the animation of the scheduling/timetabling process looks naturally. To show capabilities of the proposed algorithm we implemented a timetabling system in Java that consists of independent modules with a generic scheduling engine and with a graphical user interface.

## Motivation and Problem Description

A traditional scheduling problem is defined by a set of activities and by a set of resources to which the activities are allocated. Moreover, the activities must be positioned in time (otherwise, the problem is called resource allocation). Timetabling can be seen as a special form of the scheduling problem with slightly simplified view of time. In the timetabling problem, the time slots are defined uniformly for all the resources so the activities are being allocated to these slots (rather than to particular time).

As the basic motivation of our research we took a school timetabling problem and we extracted the basic features to model timetabling problems in general. The basic object to be scheduled in a typical school timetabling problem is a **lecture**. The scheduled time interval, typically a week, is divided into the **time slots** with the same duration. For example, we have ten slots per day where every slot takes 45 minutes. Every lecture has assigned its duration, expressed as a number of the time slots; this number is usually rather small - from one to three. Next, we have a set of resources, e.g. **classrooms** where the lectures take place, **teachers** who teach the lectures, **classes** for which the lectures are given, and other resources like overhead projector etc. Capacity of every resource is limited to one lecture. It may seem strange to consider classes as a resource but doing this allows us to define naturally constraints like no overlap of the lectures for a single class or a common lecture for several classes. Note that the set of slots is assigned to every resource so when we are allocating the lectures to the slots, it means allocation to a particular slot of a particular resource. It is possible to disable some slots in the resource to express forbidden time when no activity can be scheduled, e.g. the teacher is not available, the classroom is occupied by some external activity etc. Naturally, it is also possible to put preferences to the slots expressing to which slots the activities should be preferentially allocated.

Every lecture has assigned a teacher, a class, a set of possible classrooms (one of them is selected), and a set of special resources (overhead projector etc.). So when we are allocating a lecture, we are putting the lecture to equivalent time slots in all requested resources. Thus, the task is to choose a start time (a start lot) of every lecture as well as to select one resource among alternative resources (e.g. classrooms). Until now the relationship between the lectures is indirect only, via resources (no two different lectures can be allocated to a single slot of the same resource). Sometimes, the relation between the lectures should be explicit, for example to express that two alternative lectures are taught in parallel or that the lesson is before the exercises. We call this relation a **dependency**.

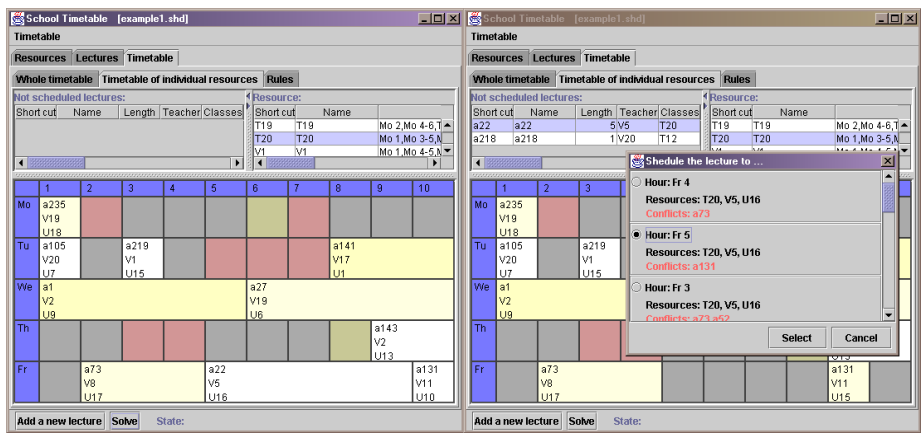

**Fig. 1.** In interactive timetabling, the user see continuous improvements of the schedule (left) and he or she can interfere with this process, e.g. by manual allocation of the activity (right).

In the following section, we propose a general model motivated by the above school timetabling problem. However, we believe that this model can be applied to other timetabling problems different from this particular school timetabling problem. Still, there is a space for other extensions of this model, like introduction of new preferences or dependencies.

## The Model

We propose a generic model for timetabling problems consisting of a set of resources, a set of activities, and a set of dependencies between the activities. Time is divided into the time slots with the same duration. Every slot may have assigned a constraint, either hard or soft one: a hard constraint indicates that the slot is forbidden for any activity, a soft constraint indicates that the slot is not preferred. We call these constraints "time preferences". Every activity and every resource may have assigned a set of time preferences, which indicate forbidden and not preferred time slots.

**Activity** (which can be directly mapped to a lecture) is identified by its name. Every activity is described by its **duration** (expressed as a number of time slots), by **time preferences**, and by a **set of resources**. This set of resources determines which resources are required by the activity. To model alternative as well as required resources, we divide the set of resources into several subsets – **resource groups**. Each group is either conjunctive or disjunctive: the conjunctive group of resources means that the activity needs all the resources from the group, the disjunctive group means that the activity needs just one of the resources (we can select among several alternatives). An example can be a lecture, which will take place in one of the possible classrooms and it will be taught for all of the selected classes. Note that we need not model conjunctive groups explicitly because we can use a set of disjunctive groups containing exactly one resource instead (the set of required resources can be described in a conjunctive normal form). However, usage of both conjunctive and disjunctive groups simplifies modelling for the users.

**Resource** is also identified by its **name** and it is fully described by **time preferences**. There is a hard restriction, that only one activity can use the resource in the same time. As we mentioned in the previous section, the resource represents a teacher, a class, a classroom or another special resource at the school timetabling problem.

Finally, we need a mechanism for defining and handling direct **dependencies** between the activities. It seems sufficient to use binary dependencies only that define relationship between two activities. Currently, we use a selection of three temporal constraints: the activity finishes before another activity, the activity finishes exactly at the time when the second activity starts, and two activities run concurrently (they have the same start time). The scheduling engine provides an interface for introduction of other binary dependencies.

**The solution** of the problem defined by the above model is a timetable, where every scheduled activity has assigned its start time and a set of reserved resources, which are needed for its execution (the activity is allocated to respective slots). This timetable must satisfy all the hard constraints, namely:
- every scheduled activity has all required resources reserved, i.e., all resources from the conjunctive groups and one resource from each disjunctive group of resources,
- two scheduled activities cannot use the same resource at the same time,
- no activity is scheduled into a time slot where the activity or some of its reserved resources has a hard constraint in the time preferences,
- all dependencies between the scheduled activities must be satisfied.

Furthermore, we want to minimise the number of violated soft constraints in the time preferences of resources and activities. We do not formally express this objective function; these preferences will be used as a guide during search for the solution.

Last but not least, in interactive timetabling we want to present some solution to the user at each time. Therefore, we will work with partial solutions where some of the activities are not scheduled yet. Still, these partial solutions must satisfy the above constraints. Note that using a partial solution allows us to provide some reasonable result even in case of over-constrained problems.

## An Interactive Solver

To satisfy needs of interactive timetabling we designed an ad-hoc algorithm for solving the above described timetabling problem. Nevertheless, we believe that this algorithm can be generalised to solve other interactive problems as well.

Let us first summarise the requirements to such interactive solving algorithm. The algorithm should provide some (partial) solution at each step. This (partial) solution must be sound (all required constraints are satisfied) and it should be good enough in respect to satisfaction of soft constraints. It is not necessary to find the optimal solution (minimal number of violated soft constraints), a good enough solution is OK. We insist more on interactivity of the algorithm that means two things: the difference between the solutions provided by the algorithm in two consecutive steps should be minimal and the user can interfere with the solution process. The user may change parameters of activities and resources, add or remove dependencies, allocate or detach any activity from the time slot. The algorithm must be able to start from such modified timetable, make it sound and continue in extending the partial solution to a complete solution.

There exist two basic approaches how to solve problems defined by means of constraints: backtracking based search that extends a partial sound solution to a complete solution and local search that decreases the number of violations in a complete solution. The idea of local search follows our requirement that there is some solution provided at each step and the difference between the solutions in consecutive steps is rather small. However, the provided solutions are not sound. On the other side backtracking based search can provide a partial sound solution at each step but it is not easy to add interactive behaviour to it. In particular, it does not support external changes of the partial solution and the difference between two consecutive solutions can be large after long backtrack.

To satisfy needs of an interactive solving algorithm we propose to mix features of above two approaches. Our algorithm uses forward based search that extends a partial sound solution to a complete solution. The next solution is derived from the previous solution by allocating some activity and removing conflicting activities from the partial schedule. Every step is guided by a heuristic that combines minimal perturbation and minimal violation of soft constraints requirements. If the user changes somehow the partial solution, the algorithm first removes all the violated activities and then it continues from a given sound solution.

**A Basic Concept of the Interactive Solver**

We propose an interactive scheduling algorithm that works in iterations. The algorithm uses two basic data structures: a set of activities that are not scheduled and a partial sound schedule. At each iteration step, the algorithm tries to improve the current partial schedule towards a complete schedule. The scheduling starts with an empty partial schedule (which is a sound schedule), i.e. all the activities are in the list of non-scheduled activities. Then it goes repeatedly from one partial solution to another sound solution until all the activities are scheduled or the maximum number of iterations is reached.

The user may interrupt the algorithm after arbitrary iteration and he or she can acquire a solution (latest or the best ever found) where perhaps not all the activities are scheduled but all the constraints on the scheduled activities are satisfied. During this interruption, the user may allocate manually some of the non-scheduled activities, weaken some constraints or make other changes (add new activities etc.) and then let the automated scheduling continue.

Look now what is going on in the iteration step. First, the algorithm selects one non-scheduled activity and computes the locations where the activity can be allocated (locations with the hard constraint are not assumed). Every location is described by a start time (the number of the first slot for the activity) and by a set of required resources. Then every location is evaluated using a heuristic function over the current partial schedule. Finally, the activity is placed to the best location that may cause some conflicts with already scheduled activities. Such conflicting activities are removed from the schedule and they are put back to the list of non-scheduled activities.

```
procedure solve(unscheduled, schedule, max_iter)
  // unscheduled is a list of non-scheduled activities
  // schedule is a partial schedule (empty at the beginning)
     iterations=0;
     while unscheduled non empty & iterations<max_iter
           & non user interruption do
             iterations ++;
             activity = selectActivityToSchedule(unscheduled);
             unscheduled -= activity;
             location = selectPlaceToSchedule(schedule, activity);
             unscheduled += place(schedule, activity, location)
                // place the activity and return violated activities
     end while;
     return schedule;
end solve
```

**Fig. 2.** A kernel of the interactive scheduling algorithm

The above algorithm schema is parameterised by two functions: activity selection and location selection. We can see these functions in a broader view of constraint programming as variable selection and value selection functions. Note that value/variable selection functions can be found both in backtracking based search as well as in local search algorithms and there exist some general techniques used to define these functions. Our usage of these functions is somewhere in between backtracking and local search. We are selecting non-scheduled activity (i.e., non-instantiated variables) but during this selection we use information about the previous locations of the activity. Remind that the activity might be removed from the schedule during some previous iteration step. Moreover, because of removing activities from the partial schedule we need some mechanism to prevent cycles. Techniques used for activity and location selection and for escaping from a cycle are described in next sections.

**Activity Selection (a variable selection criterion)**

As mentioned above, the proposed algorithm requires a function that selects a non-scheduled activity to be placed in the partial schedule. This problem is equivalent to what is known as a variable selection criterion in constraint programming. There exist several general guidelines how to select a variable. In local search, usually a variable participating in the largest number of violations is selected first. In backtracking based algorithms, a first-fail principle is often used, i.e., a variable whose instantiation is the most complicated is selected first. This could be a variable involved in the largest set of constraints (static criterion) or a variable with the smallest domain (dynamic criterion) etc. [3, 4, 7, 8, 11]

We follow the general variable selection rules and we recommend selecting the hardest to schedule activity first. For simplicity reasons, we call it the worst activity. The question is how to decide between two activities which one is the worse activity. We can use both static and dynamic information when deciding about the worst activity. An example of static information is a number of dependencies in which the activity participates. More dependencies mean worse activity. Dynamic information is information which arises from the partial schedule, for example a number of locations in the timetable that don't conflict with any already scheduled activity. It is more expensive to acquire dynamic information but this information is usually more accurate than static information. Moreover, if static information is used alone then the solving algorithm can be easily trapped in a cycle.

In our algorithm we currently use the following criteria when selecting the activity:
- How many times was the activity removed from the timetable? ($N_{\#Rm}$)
- In how many dependencies does the activity participate? ($N_{\#Dep}$)
- In how many places can this activity be placed? ($N_{\#Plc}$)
  Note: Places, where another activity lays and can't be removed (the activity was fixed by the user), are not counted.
- In how many places this activity can be placed with no conflict? ($N_{\#PlcNC}$)

A weighted-sum of above criteria is used as follows:

$$val_{activity} = -w_1 N_{\#Rm} - w_2 N_{\#Dep} + w_3 N_{\#Plc} + w_4 N_{\#PlcNC}$$

Activity with the minimal heuristic value is selected. Notice that the first two criteria are used with a negative weight, this is because a larger value means worse activity. Using such formula gives us more flexibility when tuning the system for a particular problem. Moreover, it allows studying influence of a particular criterion to efficiency of the scheduling.

It is possible to select the worst activity among the all non-scheduled activities but due to complexity of computing the heuristic value, this could be rather expensive. Therefore we propose to select a subset of non-scheduled activities randomly (we use a probability of selection 0.2) and then choose the worst activity from this subset. The results will not be much worse and we can select the activity approximately five times faster (see section Results).

**Location Selection (a value selection criterion)**

After selecting the activity we need to find a location where to place it. This problem is usually called a value selection in constraint programming. Typically, the most useful advice is to select the best-fit place [3, 4, 7, 8, 11]. So, we are looking for a place, which is the most preferred for the activity and also where the activity causes less trouble. It means that we need to find a place with minimal potential future conflicts with other activities. Note that we are not using constraint propagation explicitly in our algorithm, the power of constraint propagation is hidden in the location selection.

To find the best place for the activity we explore the space of all possible start times and for each start time we explore the possible sets of resources that the activity needs for its execution. Remind that for a single start time we can have several sets of resources (there can be alternatives among the required resources). We evaluate each such location using the following criteria:

- How many scheduled activities will be in the conflict with the current activity if it is placed to the selected location? ($N_{\#CnfAct}$)
- Will the activity remove another activity repeatedly? ($N_{\#rep}$, the number of such repeatedly removed activities)
  For every activity we can register the activity which caused its removal from the schedule in any previous iteration. By including these activities in the next decision we can avoid some short cycles.
- How many scheduled activities, which are in conflict with the location, can't be rescheduled without another conflict elsewhere? ($N_{\#ConflNoRsh}$)
- How many soft constraints are violated (both in activity and selected resources time preferences)? ($N_{\#soft}$)
- How far is the location from the location where the activity was placed before? ($N_{diffPl}$)
- How good is the location from the user point of view? ($N_{user}$)
  This is a user-defined preference for location used to model real-life problems (for example, prefer morning slots to evening slots).

Again, we use a weighted sum of above criteria to evaluate the location:

$$val_{place} = w'_1 N_{\#CnfAct} + w'_2 N_{\#rep} + w'_3 N_{\#ConflNoRsh} + w'_4 N_{\#soft} + w'_5 N_{diffPl} + w'_6 N_{user}$$

Location with the minimal heuristic value is selected. To remove cycling, it is possible to randomise this location selection procedure. For example, it is possible to select five best locations for the activity and then choose one randomly. Or, it is possible to select a set of best locations such that the heuristic value for the worst location in this group is maximally twice as large as the heuristic value of the best location. Again, the location is selected randomly from this group. This second rule inhibits randomness if there is a single very good location.

**How to Escape a Cycle?**

In the previous two sections about value and variable selection we proposed some mechanisms how to avoid cycling, but we can do more. In the current implementation of the scheduling engine we use a technique based on a tabu list [3, 6, 10].

Tabu list is a FIFO (first in first out) queue of pairs (variable, value) with a fixed length. When a new value is assigned to a variable, a new pair is inserted into the tabu list and the last pair is removed from the tabu list. Now, when we are selecting a new value for the variable X we can avoid a repeated selection of the same value for the variable (value V is not selected if the pair (X,V) is in the tabu list,). This tabu rule prevents short cycles (a cycle length corresponds to the length of the tabu list). There also exist other rules, called aspiration criteria, which can brake the above tabu rule of assignment, i.e., we can assign a value satisfying the aspiration criterion even if it is in the tabu list.

In our scheduling algorithm we adapted a slightly modified version of the tabu list. We use pairs (activity, location) in the tabu list. When we choose an activity A and before we compute a heuristic value for the location L we explore the tabu list. If the pair (A,L) appears twice in the tabu list then we do not use the location L for the activity A in this iteration. If the pair (A,L) appears exactly once in the tabu list then the location L is assumed only if this location has the minimal heuristic value (it is the best location). This is also the case when the pair (A,L) could be introduced to the tabu list for the second time (if the location L is selected). Finally, if the pair (A,L) is not in the tabu list then the location is processed as described in the previous section. To summarise it, our aspiration criterion allows us to select some location for the activity maximally two times in the period defined by the length of the tabu list. The second assignment is allowed only if the location is the best for the activity.

## Preliminary Results

We have implemented the above described scheduling algorithm in Java and added a user interface for solving school timetabling problems (see Figure 1). In this section we present some preliminary results achieved using randomly generated problems with the size 20 classes, rooms, and teachers and 10 slots per day (5 days). In these tests, we used fix weights for all the criteria described in the sections about activity and location selection. We compared the three basic activity selection criteria, namely random selection of the activity, selection of the worst activity among 20% randomly selected non-scheduled activities, and selection of the worst activity among all non-scheduled activities. Feasible problems with different number of activities were used (the number of activities is measured in % of the timetable space).

It is not surprising that the selection of the activity from all non-scheduled activities leads to the smallest number of iterations and to schedules with the largest number of scheduled activities. However, this method also brings some overhead so it needs more time to find a schedule. Random selection of the activity is very fast but it requires higher number of iterations to find a schedule and in cases with more activities the number of scheduled activities decreases. Our experiments confirm that the idea of combination of random selection with activity selection heuristic could bring interesting results. The number of iterations and successfully scheduled activities is comparable to the method where all non-scheduled activities are explored. However, the overall time is lower because we are choosing from about 20% of randomly selected activities only.

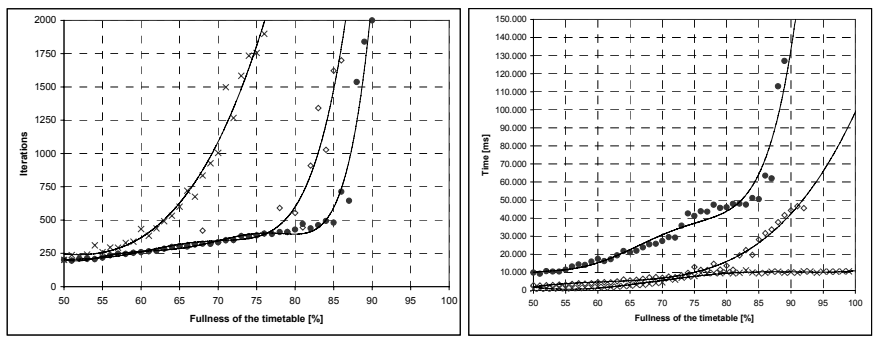

**Fig. 3.** Comparison of the number of iterations and time for three basic variable selection criteria (× randomly selected activity, ◊ the worst activity among 20% randomly selected activities, ● the worst activity among all activities).

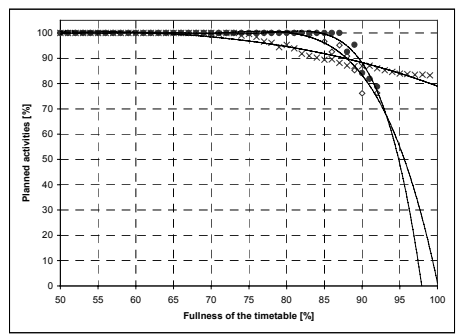

**Fig. 4.** Comparison of the number of scheduled activities for three basic variable selection criteria (× randomly selected activity, ◊ the worst activity among 20% randomly selected activities, ● the worst activity among all activities).

## Conclusions and Future Work

We presented a promising algorithm for solving timetabling problems, which combines principles of the local search and other techniques for solving constraint satisfaction problems. The presented algorithm can be used in interactive environment that is its greatest advantage. Moreover, the algorithm can be easily extended by adding more heuristics describing other soft constraints. This algorithm was proposed to solve timetabling problems with motivation in school timetabling. However, we believe that its principles can be used in other constraint satisfaction problems especially when interactive behaviour is required.

Currently, we are working on further empirical studies of this algorithm with a particular emphasis on studies how the weights influence efficiency. Further research is oriented both theoretically, to formalise the techniques and to put them in a wider context of constraint programming, and practically, to implement the engine as a system of co-operating parallel threads.